# Comparative Performance Analysis of Modern NoSQL Data Technologies: Redis, Aerospike, and Dragonfly

Deep BODRA ✉ ✉
Harrisburg University of Science and Technology, USA
https://orcid.org/0009-0009-4173-2447

Sushil KHAIRNAR ✉ ✉
Virginia Tech, USA
https://orcid.org/0009-0006-5192-0175

Abstract

The rise of distributed applications and cloud computing has created a demand for scalable, high-performance key-value storage systems. This paper presents a performance evaluation of three prominent NoSQL key-value stores: Redis, Aerospike, and Dragonfly, using the Yahoo! Cloud Serving Benchmark (YCSB) framework. We conducted extensive experiments across three distinct workload patterns (read-heavy, write-heavy), and balanced while systematically varying client concurrency from 1 to 32 clients. Our evaluation methodology captures both latency, throughput, and memory characteristics under realistic operational conditions, providing insights into the performance trade-offs and scalability behaviour of each system.

Introduction

The model digital environment involves an extraordinary level of data creation and usage, and rapid expansion of web applications, mobile technology, and Internet of Things (IoT) devices. Conventional relational database management systems, although powerful and equipped with numerous features, frequently find it challenging to satisfy the performance (Salunke & Ouda, 2024) and scalability requirements of modern applications that require sub-millisecond response times and the capacity to manage millions of operations per second. NoSQL databases, especially key-value stores, emphasize performance and scalability have been used to overcome that challenge. Key-value stores are a fundamental and commonly used NoSQL model, providing a straightforward but effective abstraction for data storage and access.

Among the key competitors in this field, Redis has positioned itself as a in-memory data store for caching, session handling, and real-time analytics. Aerospike markets itself as a high-speed, distributed database built for applications that demand both performance and reliability at scale. Dragonfly, a newer player, seeks to deliver Redis-compatible features while tackling certain scalability and performance constraints of conventional Redis implementations. Though these systems are widely used, thorough comparative evaluations that systematically evaluate their performance traits across various workload patterns and levels of concurrency are still scarce (Stjepanovic, D. et al., 2015; Anthony & Rao, 2015; Mohan, Kanmani, Ganesan & Ramasubramanian 2024).

This paper addresses this gap by presenting a comparative evaluation of Redis, Aerospike, and Dragonfly using the Yahoo! Cloud Serving Benchmark (YCSB) framework. Our study systematically examines the performance characteristics of these systems across three fundamental workload patterns: read-heavy, write-heavy, and balanced operations. By varying client concurrency from single-client scenarios to highly concurrent 32-client configurations, we capture the scalability behaviour and performance trade-offs of each system under realistic operational conditions. The contributions of this work include performance comparison including both latency and throughput metrics, analysis of memory consumption characteristics, and practical guidance for system architects in selecting appropriate key-value storage solutions.

1. Database Systems Overview

Redis is an open-source, in-memory data structure store that serves as a database, cache, and message broker (Charan, P. S. B., Varshitha, G., Lashya, A., Varma, U. S. R., & Madhusudhan, D). Redis was developed in 2009 and has become one of the most popular NoSQL databases due to its simplicity, performance, and versatility. Redis stores data in memory to enable fast read and write operations and sub-millisecond latencies. The system supports various data structures including strings, hashes, lists, sets, sorted sets, bitmaps, and streams which makes it suitable for use cases beyond simple key-value operations. Redis employs a single-threaded architecture for command processing which eliminates the need for complex locking mechanisms but can limit scalability on multi-core systems. Persistence is achieved through periodic snapshots or append-only files, providing durability options while maintaining high performance. Redis Cluster enables horizontal scaling by partitioning data across multiple nodes, but it introduces additional complexity in deployment and management (Easwaramoorthy et al., 2025; Mohan et al., 2024).

Aerospike is a distributed NoSQL database designed for high-performance applications that require speed and scale. It was found in 2009 and was built to address the limitations of traditional databases in handling real-time applications. The system uses a hybrid memory architecture that combines DRAM for index storage and SSDs for data storage to optimize performance and cost. The architecture automatically handles data partitioning, replication, and cluster management. It also offers strong consistency and automatic failover capabilities. The database utilizes a shared-nothing architecture where each node operates independently for high scalability (Volminger, A., 2021). The query engine supports both key-value operations and complex secondary index queries

Dragonfly is an in-memory data store created as a direct substitute for Redis to tackle its single-threaded scalability limitation. Launched in 2022, it utilizes a multi-threaded, shared-nothing architecture that can leverage multiple CPU cores, potentially providing much greater throughput on modern hardware. The system utilizes the Redis protocol, guaranteeing compatibility with current Redis clients and applications without the need for code modifications. Dragonfly implements sophisticated memory management methods and lock-free data structures to reduce contention and enhance performance among threads

2. Experimental Setup

All experiments were conducted on a Mac OS system equipped with an Apple M3 Pro chip featuring 12 cores and 36 GB of RAM, running macOS Sequoia. The databases were deployed locally using Docker containers to ensure consistent and isolated environments for each system. This configuration provided a controlled testing environment while leveraging the high-performance ARM architecture of the M3 Pro chip. Docker containerization enabled precise resource allocation and eliminated potential interference between different database instances during sequential testing. The substantial memory capacity and multi-core architecture of the test system allowed for a comprehensive evaluation of each database's scalability characteristics under varying concurrency levels

The evaluation utilized the Yahoo! Cloud Serving Benchmark (YCSB), a widely adopted framework for benchmarking NoSQL databases that provides standardized workloads and metrics for fair comparison across different systems. YCSB operates through a two-phase approach: the load phase populates the database with initial data, while the run phase executes the actual benchmark operations according to the specified workload characteristics. The framework's importance lies in its ability to generate realistic, configurable workloads that simulate real-world application patterns, enabling systematic performance evaluation across different operational scenarios (Ferreira et al., 2025; Beckermann, 2025). YCSB supports various data access patterns and allows precise control over concurrency levels, making it ideal for evaluating database scalability.

For this study, we configured YCSB to test with 1, 2, 4, 8, 16, and 32 concurrent clients to assess each system's performance characteristics under increasing load. The benchmark employed a Zipfian distribution for key selection, which realistically models the non-uniform access patterns commonly observed in production systems where a small subset of keys receives the majority of request.

The read-heavy workload simulated applications with predominantly read operations, configured with a 95% read and 5% update operation ratio. Each record consisted of 1 KB of data organized as 10 fields of 100 bytes each, plus the key identifier. This workload pattern is representative of caching scenarios, content delivery systems, and read-intensive web applications where data retrieval significantly outweighs modification operations. The load phase inserted 1,474,560 records to establish a substantial dataset, while the run phase performed the same number of operations with the specified read-update ratio. This configuration tests each database's ability to handle high throughput read operations while maintaining low latency under concurrent access patterns.

The balanced workload provided equal distribution of read and update operations with a 50% read and 50% update ratio, representing applications with mixed access patterns such as social media platforms, collaborative applications, and general-purpose web services. Records maintained the same 1 KB structure as the read-heavy workload, consisting of 10 fields of 100 bytes each plus the key. The load phase populated the database with 1,474,560 records, while the run phase executed the same number of operations with balanced read-update distribution. This workload evaluates each system's ability to handle concurrent read and write operations efficiently, testing both query performance and transaction processing capabilities under mixed load conditions.

The write-heavy workload utilized YCSB's time series workload template configured with a 10% read and 90% insert ratio, designed to simulate high-throughput data ingestion scenarios typical of IoT applications, monitoring systems, and real-time analytics platforms. The workload generated time series data with 64 fields per key, each field having a length of 8 characters, creating a total of 1,024 unique time series combinations. This configuration models applications that continuously ingest streaming data with occasional read operations for monitoring or alerting purposes. The load phase inserted 1,474,560 records to establish baseline data, while the run phase performed 2,949,120 insert operations, effectively doubling the dataset size. This workload tests each database's ability to sustain high write throughput while maintaining acceptable performance for concurrent read operations, evaluating both ingestion capabilities and storage efficiency under continuous data growth.

## 3. Results and Analysis

### 3.1. Read-Heavy Workload Performance

The read-heavy workload results show significant performance differences across the three database systems. Aerospike demonstrated superior performance in latency and throughput metrics, achieving the lowest P99 latencies ranging from 436ms with a single client to 2,979ms at 32 concurrent clients. This performance advantage stems from Aerospike's hybrid memory architecture, where frequently accessed data remains in DRAM while the distributed hash table enables efficient data location without centralized bottlenecks. The system's shared-nothing architecture allows each node to process requests independently, eliminating the serialization bottlenecks that plague single-threaded systems under high concurrency.

Similarly, Aerospike delivered the highest throughput, scaling from 3,348 operations per second with one client to 32,592 operations per second at maximum concurrency. Redis showed moderate performance with P99 latencies between 862ms and 4,447ms, while achieving throughput values from 1,656 to 17,158 operations per second. Redis's single-threaded event loop eliminates lock contention and ensures atomic operations but becomes the primary performance bottleneck under high concurrency. Each client request must be serialized through the main thread, creating queuing delays that manifest as increased latency at higher concurrency levels. The approximately 2.7x latency increase from 1 to 32 clients demonstrates this serialization penalty.

Dragonfly exhibited the highest latencies in this workload, ranging from 1,137ms to 4,883ms, with throughput scaling from 1,371 to 16,328 operations per second. Despite its multi-threaded design, Dragonfly's performance suggests that the coordination overhead between threads and lock-free data structure management introduces significant latency penalties for read operations. The system's attempt to maintain Redis compatibility while implementing thread-safe operations appears to create computational overhead that outweighs the benefits of parallelism in read-heavy scenarios.

Figure 1. P99 latency comparison of Redis, Aerospike, and Dragonfly

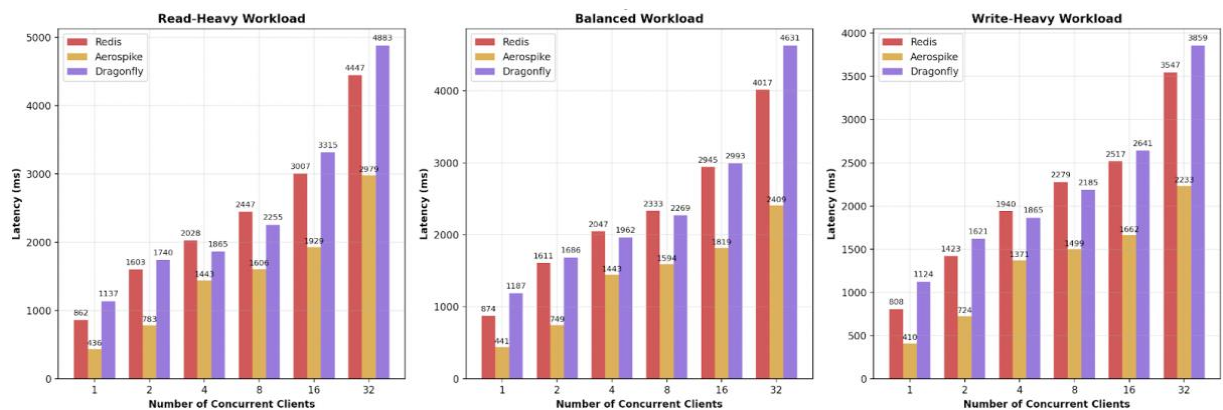

Figure 2. Throughput comparison of Redis, Aerospike, and Dragonfly

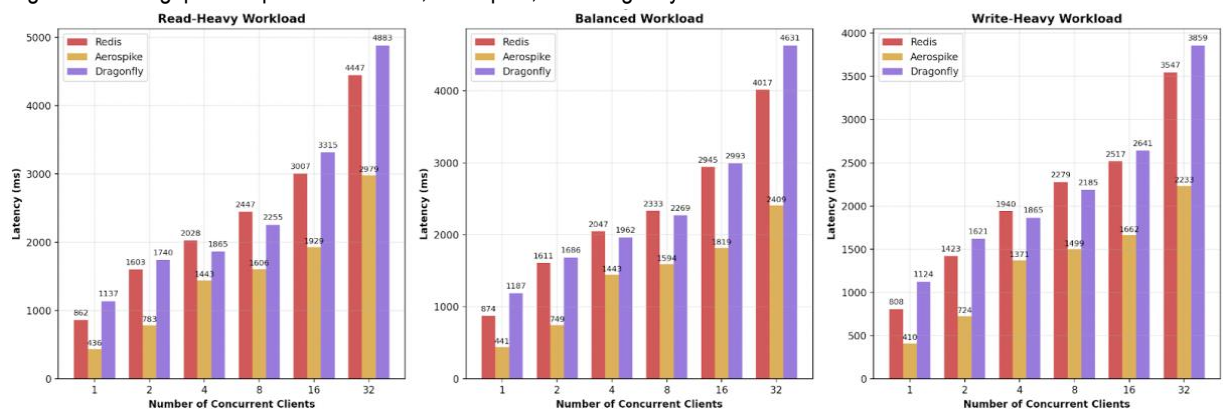

### 3.2. Balanced Workload Performance

The balanced workload showed similar performance hierarchies but with some variations. Aerospike maintained its performance leadership with P99 latencies from 441ms to 2,409ms and throughput scaling from 3,372 to 33,741 operations per second. The consistent performance across workload types demonstrates Aerospike's distributed consensus mechanisms and replication strategies effectively handle mixed read-write workloads without creating operation-specific bottlenecks.

Redis performance remained consistent with the read-heavy scenario, showing P99 latencies between 874ms and 4,017ms and throughput from 1,664 to 17,004 operations per second. This stability reflects an inherent characteristic of Redis's single-threaded architecture: operation type has minimal impact on processing efficiency since all operations are serialized through the same execution path. However, the slight latency increase in balanced workloads indicates that write operations require additional processing for persistence mechanisms (Append Only File/Redis Database) and replication.

Dragonfly showed marginal improvements compared to the read-heavy workload, with latencies ranging from 1,187ms to 4,631ms and throughput from 1,278 to 16,497 operations per second. The improvement suggests that Dragonfly's multi-threaded architecture can better distribute mixed operation types across threads, though the coordination overhead still limits overall performance gains.

### 3.3. Write-Heavy Workload Performance

The write-heavy time series workload produced the most favorable results across all systems, with generally lower latencies and higher throughput compared to read-intensive scenarios. This improvement reflects fundamental characteristics of how each architecture handles sequential write operations and reduced read-write contention.

Aerospike continued to dominate performance metrics with P99 latencies from 410ms to 2,233ms and exceptional throughput scaling from 3,562 to 34,896 operations per second. The superior write performance demonstrates Aerospike's optimized write path, where data is immediately written to memory while asynchronous background processes handle SSD persistence. The distributed architecture enables parallel write processing across nodes without coordination overhead for simple insert operations.

Redis showed its best performance in this workload with latencies ranging from 808ms to 3,547ms and throughput from 1,757 to 17,170 operations per second. The performance improvement in write-heavy scenarios reveals Redis's strength: sequential write operations benefit from the absence of lock contention and simplified memory management. The single-threaded nature becomes advantageous when operations don't require complex coordination.

Dragonfly also demonstrated improved performance with latencies between 1,124ms and 3,859ms and throughput scaling from 1,331 to 16,925 operations per second. The significant improvement in write-heavy workloads suggests that Dragonfly's multi-threaded architecture is better optimized for write operations, where thread coordination overhead is minimized and parallel processing provides tangible benefits.

### 3.4. Scalability Analysis and Architectural Implications

Examining scalability characteristics across concurrency levels reveals distinct architectural advantages and limitations. The scaling behaviour directly correlates with each system's core design philosophy and technical implementation choices.

Aerospike demonstrates near-linear throughput scaling across all workloads, increasing throughput by approximately 9-10x when scaling from 1 to 32 clients, while maintaining relatively controlled latency degradation. This scaling pattern reflects the fundamental benefit of distributed architectures: the ability to handle concurrent requests without centralized bottlenecks. Each client can potentially interact with different nodes or processing units, enabling true parallel request processing.

Redis shows consistent but more modest scalability, achieving 10-11x throughput improvements with proportionally higher latency increases, suggesting bottlenecks in its single-threaded architecture under high concurrency. The scaling limitation becomes apparent as concurrency increases beyond the system's ability to process requests through a single thread efficiently. The latency degradation follows a predictable pattern: as the request queue grows, each subsequent request experiences longer wait times. Beyond 16 concurrent clients, Redis shows signs of saturation where additional concurrency provides diminishing throughput returns while significantly increasing latency.

Dragonfly exhibits strong scalability potential with 12-13x throughput improvements from single to maximum concurrency, though starting from lower baseline performance. While the throughput scaling appears impressive, the consistently higher baseline latencies suggest that Dragonfly's multi-threaded coordination mechanisms introduce fixed overhead costs. The system appears to trade single-request efficiency for improved concurrent processing capability.

### 3.5. Cross Workload Performance Comparison and Technical Trade-offs

Analysing performance variations across different workload types reveals fundamental architectural characteristics and technical trade-offs inherent in each system's design. The workload-specific performance variations directly reflect how each system's core design decisions impact different operation patterns.

All three databases performed best under write-heavy conditions, with Aerospike showing the least performance variation across workload types, indicating robust architectural design. This consistency demonstrates that distributed architectures can maintain performance characteristics across varied workloads because they avoid single points of contention. The hybrid storage model ensures that write operations don't interfere with read performance, maintaining balanced resource utilization.

Redis demonstrated consistent behavior across all workloads with slight performance improvements in write-heavy scenarios, reflecting its optimized memory management for sequential operations. While Redis's single-threaded nature limits peak performance, it provides predictable behavior across workload types. The slight improvement in write scenarios reflects the absence of read-write coordination overhead, but the limited scalability represents a fundamental architectural constraint for high-concurrency applications.

Dragonfly showed the most significant performance variation, with better results in write-heavy workloads compared to read-intensive scenarios. This variation suggests that Dragonfly's multi-threaded implementation has uneven optimization across operation types. The better write performance indicates successful parallel write processing, while the read performance suggests that thread coordination overhead disproportionately affects read operations, possibly due to cache coherence costs or lock-free data structure complexity.

Table 1. Memory consumption comparison of Redis, Aerospike, and Dragonfly

|  | Redis (MB) | Aerospike (MB) | Dragonfly (MB) |
| --- | --- | --- | --- |
| Before the run | 36.32 | 232.1 | 58.98 |
| After the run | 2610 | 772.3 | 2350 |

*Source*: Personal research

### 3.6. Memory Consumption and Architectural Efficiency

Memory consumption patterns reveal significant differences in how each database manages storage and represents data structures. The memory utilization patterns provide insight into each system's data representation efficiency and storage optimization strategies.

Before workload execution, Aerospike demonstrated the highest baseline memory usage at 232.1 MB, reflecting its distributed architecture and metadata overhead. This initial memory footprint includes cluster management metadata, distributed hash tables, and replication state information necessary for distributed operation. While higher initially, this overhead enables the system's superior scaling characteristics.

Redis consumed 36.32 MB and Dragonfly used 58.98 MB in their initial states, representing minimal overhead single-node configurations. After completing the benchmark runs with 1,474,560 records, the memory consumption patterns shifted dramatically, revealing fundamental architectural differences.

Redis exhibited the highest memory usage at 2,610 MB, representing a 72x increase from baseline. This substantial memory overhead reflects Redis's approach of maintaining all data structures in memory with additional overhead for object metadata, expiration tracking, and internal data structure pointers. The single-threaded architecture, while simple, doesn't optimize for memory efficiency at scale.

Dragonfly consumed 2,350 MB (40x increase), suggesting that its multi-threaded architecture includes additional memory overhead for thread safety mechanisms, lock-free data structures, and coordination metadata. The improved memory efficiency compared to Redis indicates some optimization, but the thread coordination structures still impose significant overhead costs.

Aerospike used 772.3 MB (3.3x increase), demonstrating superior memory efficiency through its hybrid storage model. This efficiency stems from Aerospike's ability to store data on SSDs while maintaining only indexes and frequently accessed data in memory. The distributed architecture enables this optimization by providing fast access to persistent storage across multiple nodes.

The memory efficiency results highlight a fundamental trade-off in NoSQL database design: systems optimized for simplicity and single-node performance (Redis, Dragonfly) sacrifice memory efficiency, while distributed systems (Aerospike) achieve better resource utilization through architectural complexity.

## Conclusion

This evaluation of Redis, Aerospike, and Dragonfly using YCSB benchmarks reveals clear performance hierarchies and fundamental architectural trade-offs across diverse workload patterns and concurrency levels. Aerospike emerged as the consistent leader, demonstrating superior throughput scaling and maintaining relatively low latencies even under high concurrency, with nearly 10x throughput improvements when scaling from single to 32 concurrent clients. Redis showed stable and predictable performance across all workload patterns, though with more modest scalability due to its single-threaded architecture. Dragonfly, despite its modern multi-threaded design and Redis compatibility, consistently exhibited higher latencies compared to both competitors, though it demonstrated strong throughput scaling potential in write-heavy scenarios.

The evaluation reveals that workload characteristics significantly influence system performance, with all databases performing optimally under write-heavy conditions. These findings provide practical guidance for system selection: organizations requiring maximum performance and scalability should consider Aerospike, while those prioritizing operational simplicity may find Redis sufficient. Dragonfly represents an interesting option for Redis-compatible deployments seeking improved concurrency handling, though careful evaluation is recommended for latency-sensitive applications.

Future work should examine these systems under production-like conditions (Malyi & Serdyuk, 2024) with realistic data distributions (Abubakar et al., 2024) and in clustered mode to assess the true distributed scalability potential. Such research should address critical technical challenges including data consistency models under network partitions (particularly CAP theorem implications for each system's design), distributed transaction management across nodes, cross-datacentre replication latencies, and fault tolerance mechanisms during node failures. Additionally, investigation into network-induced performance bottlenecks, cluster coordination overhead, and the impact of different consensus algorithms on write performance would provide valuable insights into real-world deployment considerations for modern NoSQL data technologies.

## Credit Authorship Contribution Statement

Bodra, Deep led the conceptualization, methodology design, experimental implementation and analysis for database benchmarking. Khairnar, Sushil contributed to the experimental implementation, data analysis, and interpretation. Both authors reviewed and approved the final version of manuscript for publication

## Conflict of Interest Statement

The authors declare that the research was conducted in the absence of any commercial or financial relationships that could be construed as a potential conflict of interest.